\documentclass[aps,prl,twocolumn,showpacs]{revtex4-1}
\pdfoutput=1
\usepackage{epsfig}
\usepackage{graphicx}
\usepackage{bm}
\usepackage{amssymb}
\usepackage{amsmath}
\usepackage{amsthm}
\begin{document}
\title{Avalanches in 2D Dislocation Systems: Plastic Yielding is not Depinning}
\author{P\'eter Dus\'an Isp\'anovity$^1$, Lasse Laurson$^2$, Michael 
Zaiser$^3$, Istv\'an Groma$^1$, Stefano Zapperi$^{4,5}$, and Mikko J. 
Alava$^2$}
\affiliation{$^1$Department of Materials Physics, E\"otv\"os University 
Budapest, H-1517 Budapest, POB 32, Hungary}
\affiliation{$^2$COMP Centre of Excellence, Department of Applied Physics, 
Aalto University, P.O. Box 14100, FIN-00076 Aalto, Espoo, Finland}
\affiliation{$^3$Institute of Materials Simulation, Department of 
Materials Science, University of Erlangen-N\"urnberg, Dr.-Mack-Str. 77, 
90762 F\"urth, Germany}
\affiliation{$^4$CNR-IENI, Via R. Cozzi 53, 20125 Milano, Italy}
\affiliation{$^5$ISI Foundation, Via Alassio 11/C, 10126 Torino, 
Italy}

\begin{abstract}
We study the properties of strain bursts (dislocation avalanches) occurring in two-dimensional
discrete dislocation dynamics models under quasistatic stress-controlled loading. Contrary to 
previous suggestions, the avalanche statistics differs fundamentally from predictions obtained
for the depinning of elastic manifolds in quenched random media. Instead, we find an exponent 
$\tau =1$ of the power-law distribution of slip or released energy, with a cut-off that 
increases exponentially with the applied stress and diverges with system size at all stresses.
These observations demonstrate that the avalanche dynamics of 2D dislocation systems is scale-free 
at every applied stress and, therefore, can not be envisaged in terms of critical behavior 
associated with a depinning transition.
\end{abstract}
\pacs{81.40.Lm, 61.72.Lk, 68.35.Rh, 81.05.Kf}
\maketitle

Crystalline solids subject to an increasing stress undergo a transition (``yielding'') 
from nearly-elastic behavior to plastic flow by collective dislocation motion. Both during the 
run-up to yielding and in the subsequent plastic flow regime, dislocation systems exhibit 
strongly intermittent, avalanche-like dynamics. In micron sized specimens these avalanches 
show as abrupt strain bursts with a broad, power law-type size distribution \cite{DIM-06,CSI-07} 
(for a review see \cite{UCH-09}) and in larger samples they manifest themselves through acoustic 
emission (AE) events with power-law distributed amplitudes \cite{WEI-97,MIG-01}. 

Several researchers have advanced the idea that the dislocations in a crystal deforming 
under stress might be envisaged as a driven non-equilibrium system, where power-law distributed 
avalanches arise from dynamic critical behavior associated with a non-equilibrium phase transition 
at a critical value $\sigma_\text{ext} = \sigma_\text{c}$ of the externally applied stress, analogous 
to the depinning transition of elastic interfaces in random media \cite{FIS-98}. This idea applies in 
a straightforward manner to single dislocations interacting with immobile impurities which provide a 
textbook example of one-dimensional elastic manifolds undergoing a depinning transition 
\cite{ZAP-01,BAK-08}. In generalization of this observation, several authors have argued that the 
mean-field limit of the depinning transition might correctly describe the dynamic behavior of 
stress-driven many-dislocation systems even when other defects (such as impurities) are absent 
\cite{ZAI-05,ZAI-06,DAH-09, FRI-12, DER-13, TSE-13}. In this picture plastic yielding is envisaged 
as a continuous phase transition where the external stress acts as control parameter and a critical 
point is reached at the yield stress. There are several motivations for such an analogy: 
(i) dislocation-dislocation interactions are of long-range nature, implying that a mean-field 
description could be applicable, and (ii) the strain burst distribution appears to be a power law, 
$P(\Delta \gamma) \propto \Delta \gamma^{-\tau}$, with $\tau$ found to be 
$\tau \approx 1.5$ both experimentally and numerically \cite{DIM-06,CSI-07,NG-08,BRI-08,ZAI-08}, 
in apparent agreement with mean-field depinning (MFD) \cite{FIS-98}.

\begin{figure}
\includegraphics[angle=0,scale=0.29]{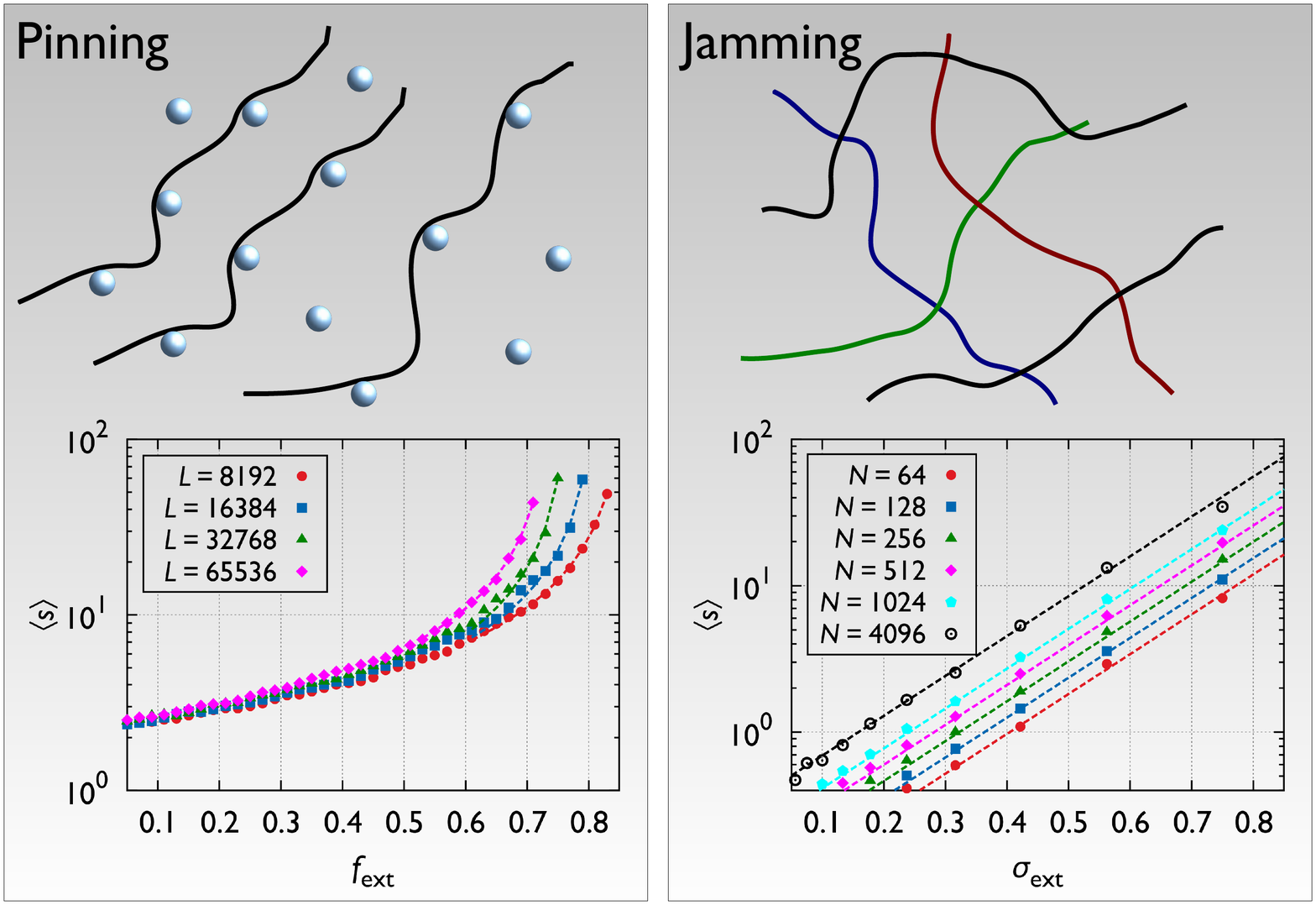}
\caption{(color online) Differences between pinning and the present jamming/unjamming 
scenario. Pinning is induced by quenched disorder which stops the motion 
of driven elastic manifolds for applied forces $f_\text{ext}$ below a critical value $f_c$ (top 
left). With $f_\text{ext}$ approaching $f_c$ from below, the manifold moves ahead in avalanches
with an average avalanche size $\langle s \rangle$ which in MFD diverges as $\langle s \rangle 
\propto (f_{c}(L)-f_\text{ext})^{-1}$ (1$d$ elastic manifold with elastic interactions
decaying as $1/r$, bottom left). $f_c$ depends on the system size $L$ due to
finite size scaling, $f_c(L) = f_c(\infty) + aL^{-1}$. In a dislocation system without 
quenched disorder, dislocation motion may stop due to formation of various jammed 
dislocation configurations (top right). The behavior of $\langle s \rangle$ we observe 
in this case is fundamentally different from the depinning scenario, with 
$\langle s \rangle = A(N) e^{\sigma_\text{ext}/\sigma_0}$, where the 
prefactor $A(N)$ grows with the number of dislocations $N$ (bottom right).}
\label{fig:illustration}
\end{figure}

There are, however, several unresolved issues regarding the validity
of the depinning picture. In the classical depinning scenario, an elastic 
manifold interacts with a \emph{static} (quenched) pinning field representing 
immobile impurities of the medium. However, yielding and avalanche dynamics of plastic
flow are generic features of crystal plasticity which do not require impurities 
or other types of quenched disorder.  Discrete dislocation dynamics (DDD) models 
\cite{MIG-02,MAD-02,WEY-02,ARS-07,ELA-08,KER-12}, which are commonly used to model 
plasticity of pure fcc crystals, relate the yield stress to the mutual trapping 
(or {\it jamming} \cite{MIG-02,LAU-10}) which occurs as interacting dislocations 
form complex metastable structures even in the absence of other defects 
\cite{CSI-07,MIG-01,MIG-02,BUL-06}. This important difference is illustrated schematically in 
Fig.~\ref{fig:illustration}.

Even if we consider the dynamics of the simplest possible DDD model -- a 2$d$ 
system of straight parallel dislocations moving on a single slip system -- there
are several findings which are not consistent with MFD. These include: (i) 
For the relaxation exponent of the Andrade creep law, i.e., the initial power 
law decay of the mean strain rate under constant applied stress, $\langle \dot{\gamma} (t)\rangle  \propto 
t^{-\theta}$, one finds the value $\theta \approx 2/3$ \cite{MIG-02, ISP-11} 
whereas MFD predicts $\theta=1$ for the critical relaxation of the order 
parameter \cite{FIS-98,MOR-04}. Also the temporal scaling of the spatial fluctuations of the 
local creep rates indicates non-mean field behaviour \cite{ROS-10}. Moreover, the duration 
of the power law relaxation regime is at low stresses limited by the system size rather 
than by the distance $\sigma_\text{ext}-\sigma_\text{c}$ from the critical point \cite{ISP-11}, 
again inconsistent with interface depinning. (ii) The stress-dependence of the 
steady state strain rate obeys $\langle \dot{\gamma} \rangle \propto (\sigma_\text{ext}-\sigma_c)^{\beta}$, 
with $\beta \approx 1.8$ \cite{MIG-02}, while MFD predicts $\beta=1$ \cite{FIS-98}. 
(iii) The response of the $2d$ DDD model to cyclic applied stresses 
is not consistent with MFD \cite{LAU-12}.  

Finally, we note that comparisons between theoretical and experimental values of 
avalanche exponents might be misleading, since most studies of dislocation avalanche 
statistics consider aggregate distributions (integrated over the different $\sigma_\text{ext}$-values), 
whereas the theoretical MFD prediction $\tau = 1.5$ refers to stress-resolved 
distributions. It is known that averaging the distributions over $\sigma_\text{ext}$ 
yields an exponent which is larger than the one of stress-resolved distributions 
\cite{DUR-06}, so the numerical and experimental findings of $\tau \approx 1.5$ by
themselves do not provide strong evidence for the MFD scenario.

In this Letter, we report results of quasistatic stress-controlled simulations of 
$2d$ DDD models and demonstrate that the stress-resolved avalanche statistics does 
not follow the MFD predictions.  To underline the general nature of our findings, 
we consider a continuous time dynamics (CTD) model with continuous spatial 
resolution and a linear force-velocity relation for the dislocations, together with 
two spatially discrete cellular automaton (CA) models with different dynamics, 
finding the same results in all cases.

The DDD models considered here are minimal representations of dislocation systems
in deforming crystals, consisting of straight parallel edge dislocations moving on
a single slip system. This implies that the problem reduces to the dynamics of $2d$ 
systems of point-like objects (the intersection points of the straight parallel 
dislocation lines with a perpendicular plane) which move on parallel lines in the $x$ 
direction of a 2$d$ Cartesian coordinate system. We consider simulation areas of size 
$L \times L$ containing $N$ dislocations with Burgers vectors ${\bf b}_i = s_i(b,0)$ 
where $s_i \in \{1,-1\}$ and $i \in [1 \, .. \, N]$. We assume equal numbers of dislocations 
of positive and negative signs. The CTD equations of motion read
\begin{equation}
\dot{x}_i = MF_i = Mb s_i \left[ \sum_{j\neq i} s_j \sigma_\text{ind} ({\bf r}_i - {\bf r}_j)
+ \sigma_\text{ext} \right], \ \ \dot{y}_i = 0.
\label{eq:eom}
\end{equation} 
Here $M$ is the dislocation mobility, $F_i$ the $x$ component of the force per unit 
length acting on the $i$th dislocation, $\sigma_\text{ind}({\bf r}) = Gb \cos (\phi) 
\cos (2\phi) r^{-1}$ is the shear stress field generated by an individual dislocation 
(with periodic boundary conditions assumed in both $x$ and $y$ directions, for details 
see \cite{BAK-06}), $G$ an appropriate elastic constant, and $\sigma_\text{ext}$ is 
the external shear stress. The CA models are defined by discretizing the system both in 
space and time. Dislocations are allowed to move from one cell to a neighboring cell if 
such a move decreases the elastic energy of the system. We apply two different rules for the 
dynamics: (i) In extremal dynamics (ED) at each step only the move which produces the largest
energy decrease is carried out. (ii) In random dynamics (RD) the moved dislocation is selected 
randomly from those that are allowed to move. The motivation of using all these models together 
is twofold: In the CA models it is easier to collect large amounts of statistics from larger 
systems, and the two CA dynamics correspond to highly non-linear relations between the acting 
force and the mean velocity of a dislocation. This makes it possible to test the generality 
of our results by comparing them with the linear force-velocity characteristics of the 
CTD model, Eq. (\ref{eq:eom}). In what follows, we measure lengths, times and stresses in 
units of $\rho^{-0.5}$, $(\rho M G b^2)^{-1}$, and $G b \rho^{0.5}$, with $\rho=N/L^2$ the 
dislocation density \cite{CSI-09}. As an example, it is noted that in these dimensionless 
units $N=L^2$.

A quasistatic stress-controlled loading protocol is implemented as follows. 
First, a random initial dislocation configuration is let to relax at 
$\sigma_\text{ext}=0$ into a metastable arrangement. Then, for the continuous 
time model, $\sigma_\text{ext}$ is increased at a slow rate from zero until 
the average dislocation velocity $V(t) = (1/N) \sum_i |\dot{x}_i(t)|$ exceeds a 
small threshold value $V_\text{th}$. While $V(t)>V_\text{th}$ and an avalanche 
propagates, the external stress is kept constant, and the amount of slip 
$s = \sum_i s_i \Delta x_i$ and plastic strain $\Delta \gamma = 
s/L^2$ produced within the avalanche are recorded. Here $\Delta x_i$ 
denotes the displacement of the $i$th dislocation during the given 
avalanche. After the avalanche is finished ($V(t) < V_\text{th}$), the external 
stress is again increased at a slow rate until the next avalanche is triggered. 
A similar loading protocol is implemented in the CA models: In between the 
avalanches, the external stress is increased just enough to make exactly one 
dislocation move, which then may trigger further dislocation activity, during 
which the applied stress is again kept constant.

\begin{figure*}
\begin{center}
\hspace*{0.0cm}
\includegraphics[scale=0.48, angle=-90]{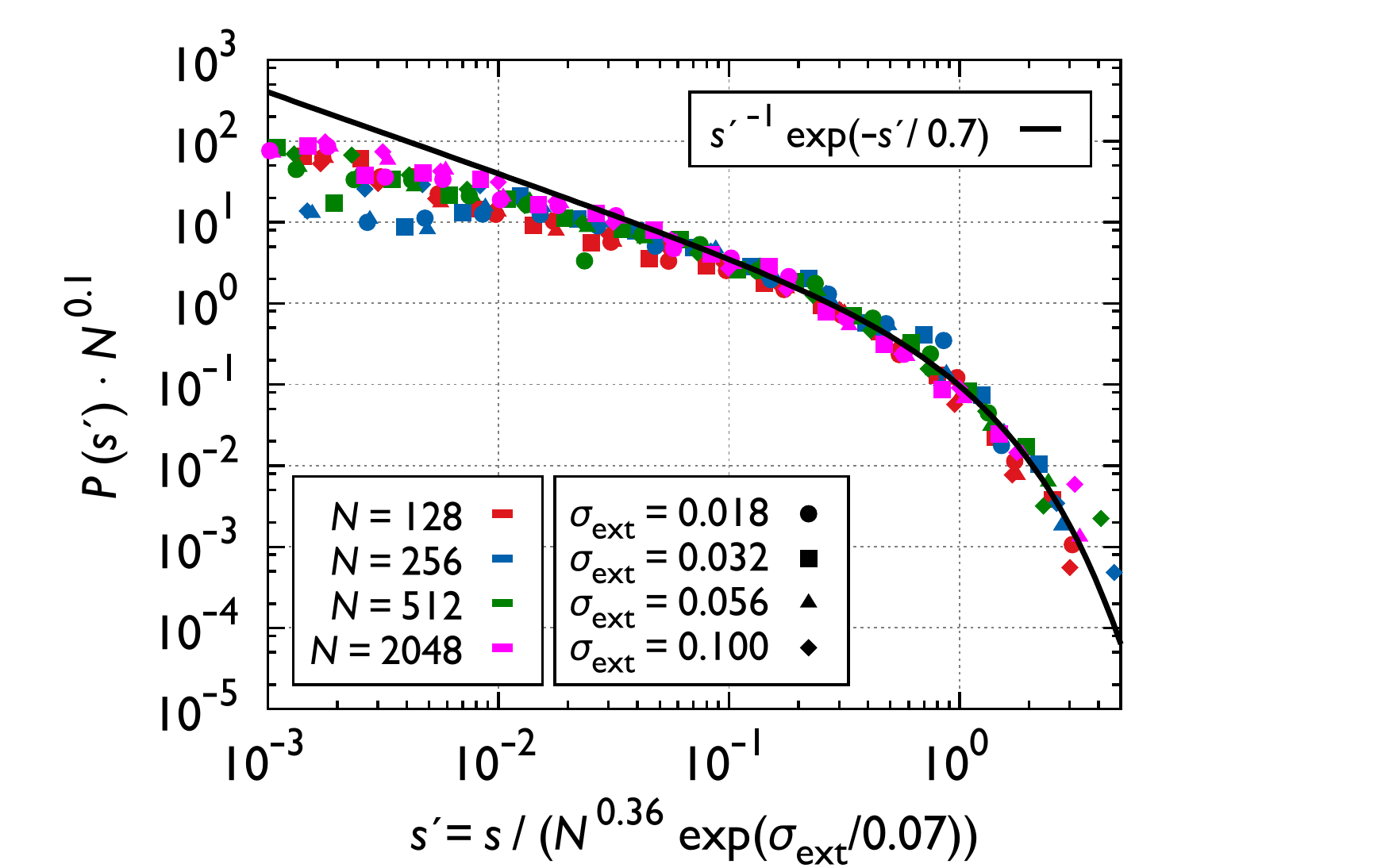}
\begin{picture}(0,0)
\put(-235,-14){\sffamily{(a)}}
\end{picture}
\hspace*{0.3cm}
\includegraphics[scale=0.48, angle=-90]{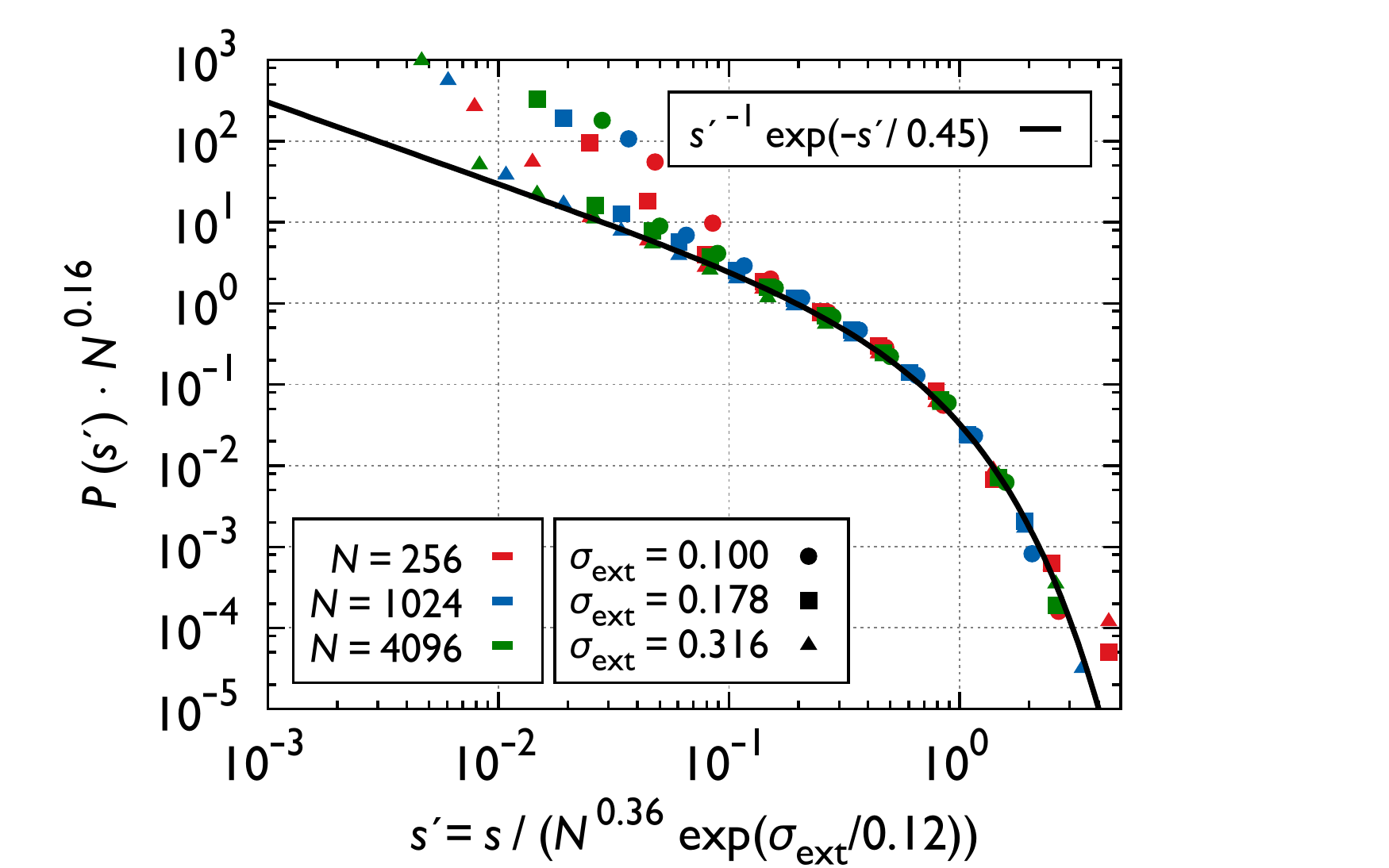}
\begin{picture}(0,0)
\put(-235,-14){\sffamily{(b)}}
\end{picture}\\
\hspace*{0.0cm}
\includegraphics[scale=0.48, angle=-90]{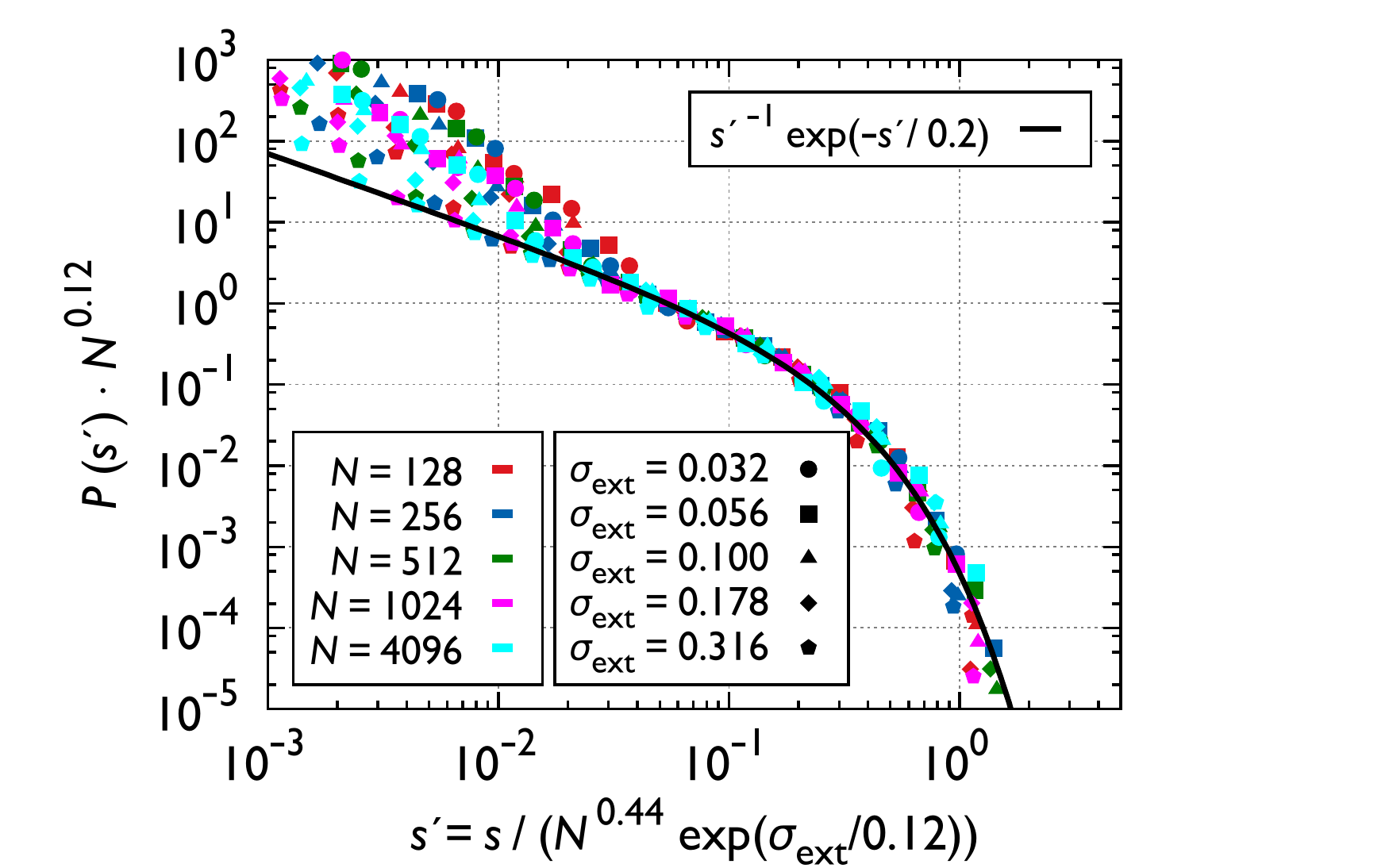}
\begin{picture}(0,0)
\put(-235,-14){\sffamily{(c)}}
\end{picture}
\hspace*{0.3cm}
\includegraphics[scale=0.48, angle=-90]{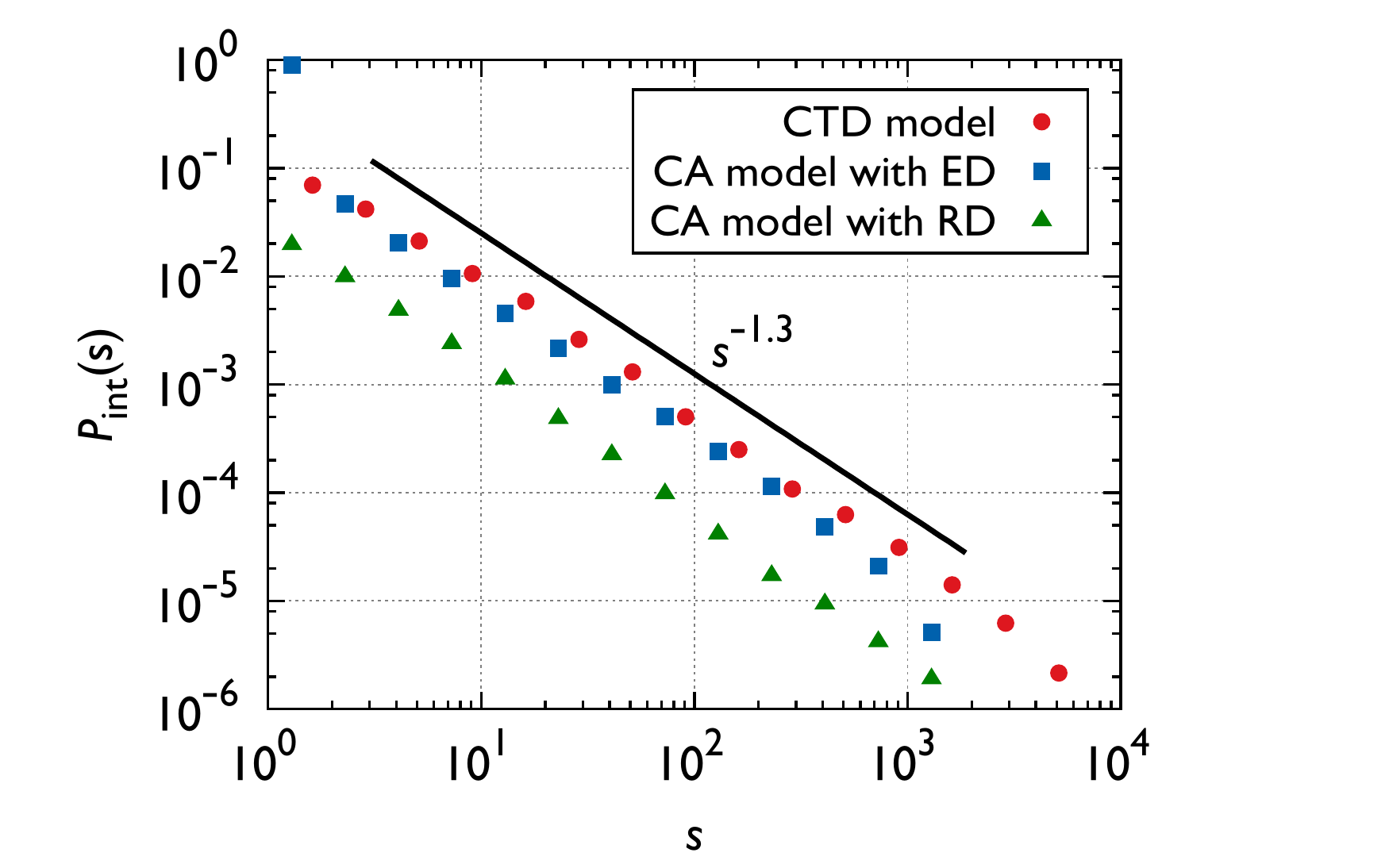}
\begin{picture}(0,0)
\put(-235,-14){\sffamily{(d)}}
\end{picture}\\
\end{center}
\caption{(color online) (a-c) Stress-resolved distributions of avalanche sizes for the various DDD 
models at different applied stresses and system sizes. The distributions are plotted as 
functions of $s' := s/s_0$, with $s_0$ obeying Eq.~(\ref{eq:cutoff_scaling}). (a) CTD model, (b) 
CA model with ED, (c) CA model with RD. (d) Aggregate avalanche size distributions $P_\text{int}$ 
integrated over $\sigma_\text{ext}$ for system sizes $N=512$ (CDT model) and $N=4096$ (CA models). 
\label{fig:avalanche_distrib}}
\end{figure*}
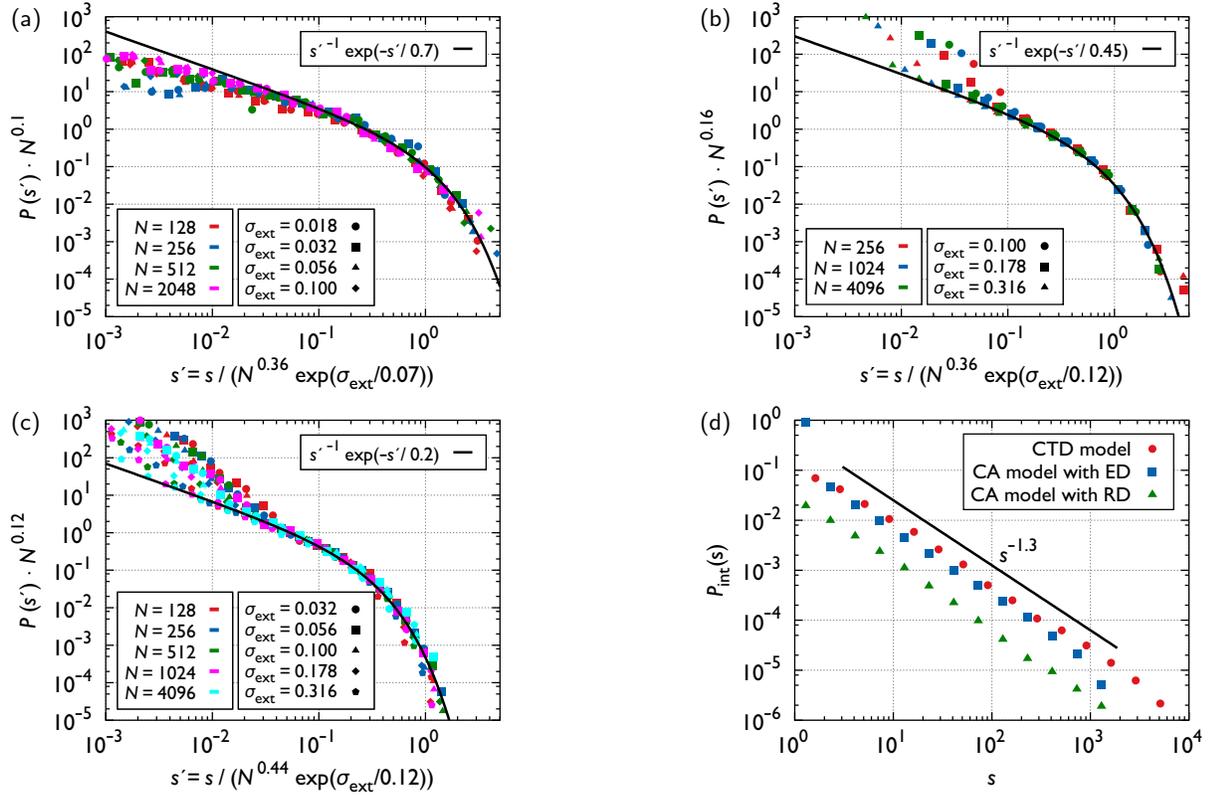

For each model, we consider the avalanche size distribution $P(s)$ 
at different levels of the external stress below the yield stress. 
For $s > 1$ (i.e. slip events larger than that corresponding to a 
single dislocation moving one average dislocation spacing), these can be well 
characterized by a power law with a cut-off,
\begin{equation}
P(s) \propto s^{-\tau} f(s/s_0).
\label{eq:av_distrib}
\end{equation}
To estimate $\tau$ and $s_0$, a fitting procedure has been used
that fits Eq.~(\ref{eq:av_distrib}) simultaneously to the avalanche distributions 
obtained at different stress levels and system sizes. The cutoff was found to follow
\begin{equation}
s_0(\sigma_\text{ext}, N) \propto N^{\beta} \exp(\sigma_\text{ext} / \sigma_0).
\label{eq:cutoff_scaling}
\end{equation}
Table \ref{tab:exponents} compiles the parameters obtained by fitting Eqs.~(\ref{eq:av_distrib}) 
and (\ref{eq:cutoff_scaling}) to the avalanche size distributions. 
Fig.~\ref{fig:avalanche_distrib}(a-c) shows the $P(s)$ distributions for the three models
plotted as functions of $s / s_0$. The validity of Eq.~(\ref{eq:cutoff_scaling}) is
demonstrated by the collapse of all distributions in the cutoff region. Since 
Eq.~(\ref{eq:av_distrib}) holds only for $s>1$, the curves follow the master curve
only as long as $s / s_0 > 1/ s_0$, thus over longer range as the applied stress 
and/or the system size increase. Below this regime the behavior is governed
by the single-dislocation dynamics and therefore differs between the three models. 

\begin{figure}
\begin{center}
\hspace*{0.0cm}
\includegraphics[scale=0.48, angle=-90]{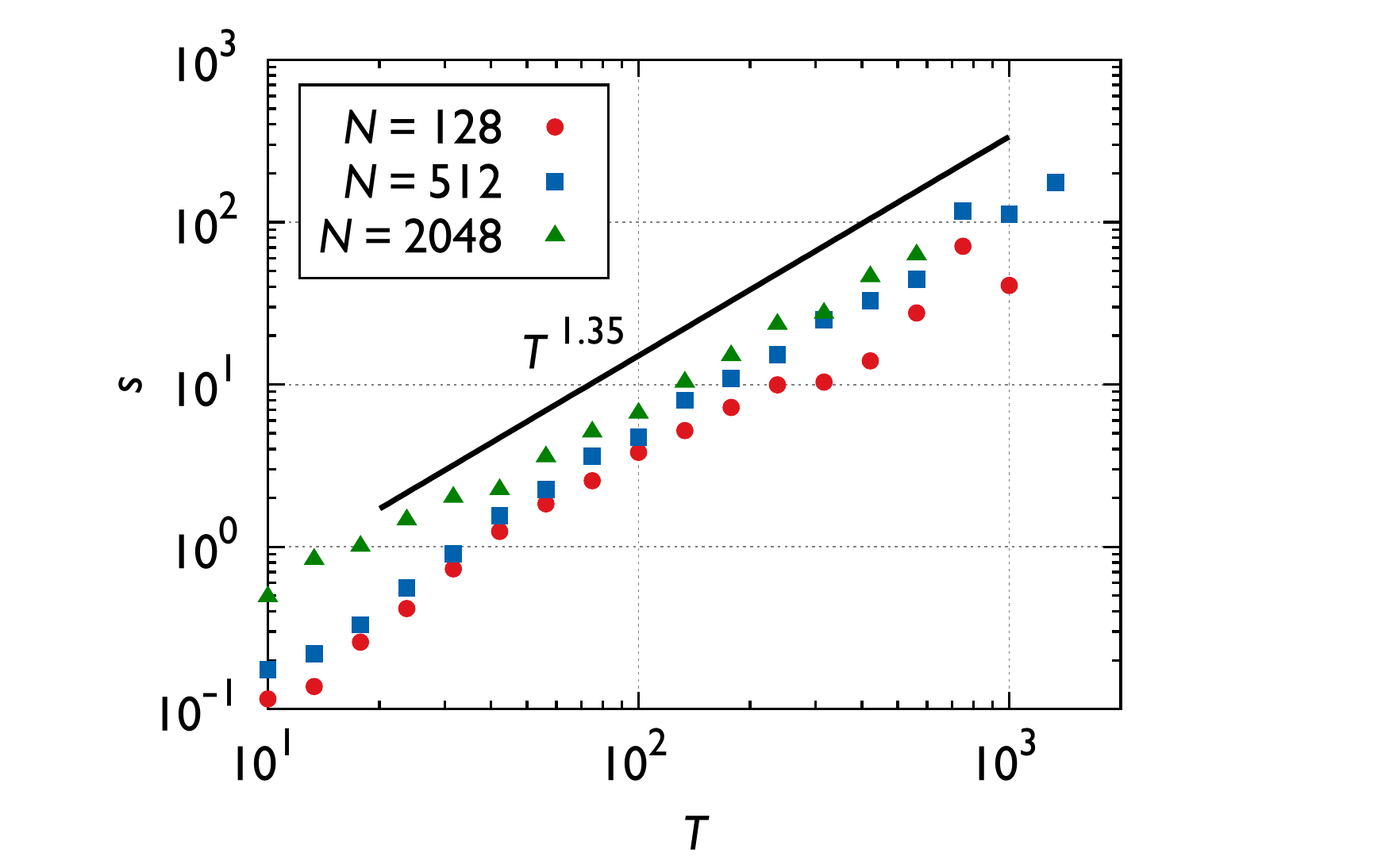}
\end{center}
\caption{
(color online) Scaling of the avalanche size $s$ with the duration $T$ in the CTD model,
for three different system sizes $N$. Notice how $s$ for a given $T$ increases with $N$.}
\label{fig:svsT}
\end{figure}

The observations summarized by Eqs.~(\ref{eq:av_distrib}), 
(\ref{eq:cutoff_scaling}) and Table \ref{tab:exponents} exhibit several interesting features
which are the main results of this Letter: 
\begin{itemize}
\item[(i)] The power law exponent $\tau$ has the value 
$\tau \approx 1.0$, clearly different from the MFD value $\tau=1.5$. According to Fig.~\ref{fig:avalanche_distrib}(d), the integrated 
distribution (where avalanches with all stress values are considered together) exhibits a 
larger exponent $\tau_\text{int} \approx 1.3$, which is in line with a recent reanalysis 
of experimental micropillar compression data \cite{ZHA-12}. Moreover, Fig.~\ref{fig:svsT}
shows that in the CTD model, the avalanche size scales with the duration as $s \propto T^{\gamma}$,
with $\gamma \approx 1.35$ clearly different from the MFD value of 2.
\item[(ii)] According to Eq.~(\ref{eq:cutoff_scaling}), the cutoff $s_0$ 
increases with system size even at very small applied stress like $s_0 \propto N^\beta$ 
with $\beta \approx 0.4$.
\item[(iii)] The cutoff $s_0$ does not diverge at a certain external stress, rather it exhibits an exponential stress dependence.
\end{itemize}
The fundamental difference between the present and depinning behavior is highlighted in Fig.~\ref{fig:illustration}, where the average avalanche size is compared for a simple model showing depinning behavior and the CA DDD model with ED considered here.

\begin{table}
\caption{Parameters of Eqs.~(\ref{eq:av_distrib}) and (\ref{eq:cutoff_scaling}) obtained by 
fitting to the numerically obtained avalanche distributions. 
\label{tab:exponents}}
\begin{ruledtabular}
\begin{tabular}{cccc}
Model & $\tau$ & $\beta$ & $\sigma_0$ \\
\hline
CTD & $0.97 \pm 0.03$ & $0.36 \pm 0.04$ & $0.07 \pm 0.01$ \\
CA with ED & $1.00 \pm 0.03$ & $0.36 \pm 0.02$ & $0.116 \pm 0.004$ \\
CA with RD & $1.02 \pm 0.01$ & $0.44 \pm 0.01$ & $0.122 \pm 0.002$
\end{tabular}
\end{ruledtabular}
\end{table}

Equation (\ref{eq:cutoff_scaling}) implies that the cutoff of the 
plastic strain bursts $\Delta \gamma$ scales as $\Delta \gamma_0 = s_0/L^2 
\propto N^{\beta-1} = L^{2\beta-2}$. Since $\beta < 1$, with increasing system size the 
observed plastic strain events get smaller, in line with the experimental evidence that 
macroscopic plasticity is a smooth process. We note that a similar scaling form (with 
$\beta \approx 0.5$) has been proposed for systems deforming in the strain hardening regime
above the yield stress \cite{CSI-07,ZAI-07}. The remarkable new finding here is that the 
same scaling holds also for very small stresses far below the yielding threshold. 
Furthermore, the 
energy dissipated during an avalanche (at a given external stress) scales as 
$E \propto \sigma_\text{ext} \Delta \gamma L^2 = \sigma_\text{ext} s$ 
\cite{ZAI-05}, i.e. it diverges for large specimens at any  
applied stress. This is in accordance with AE results obtained during 
creep experiments on large ice single crystals which show that even for resolved shear 
stresses far below the yield stress, the energy releases recorded during AE events 
exhibit a power-law distribution which spans more than six orders of magnitude without 
any apparent cutoff \cite{MIG-01,WEI-97}. 

\begin{figure}
\begin{center}
\vspace*{-0.6cm}
\includegraphics[scale=0.48, angle=-90]{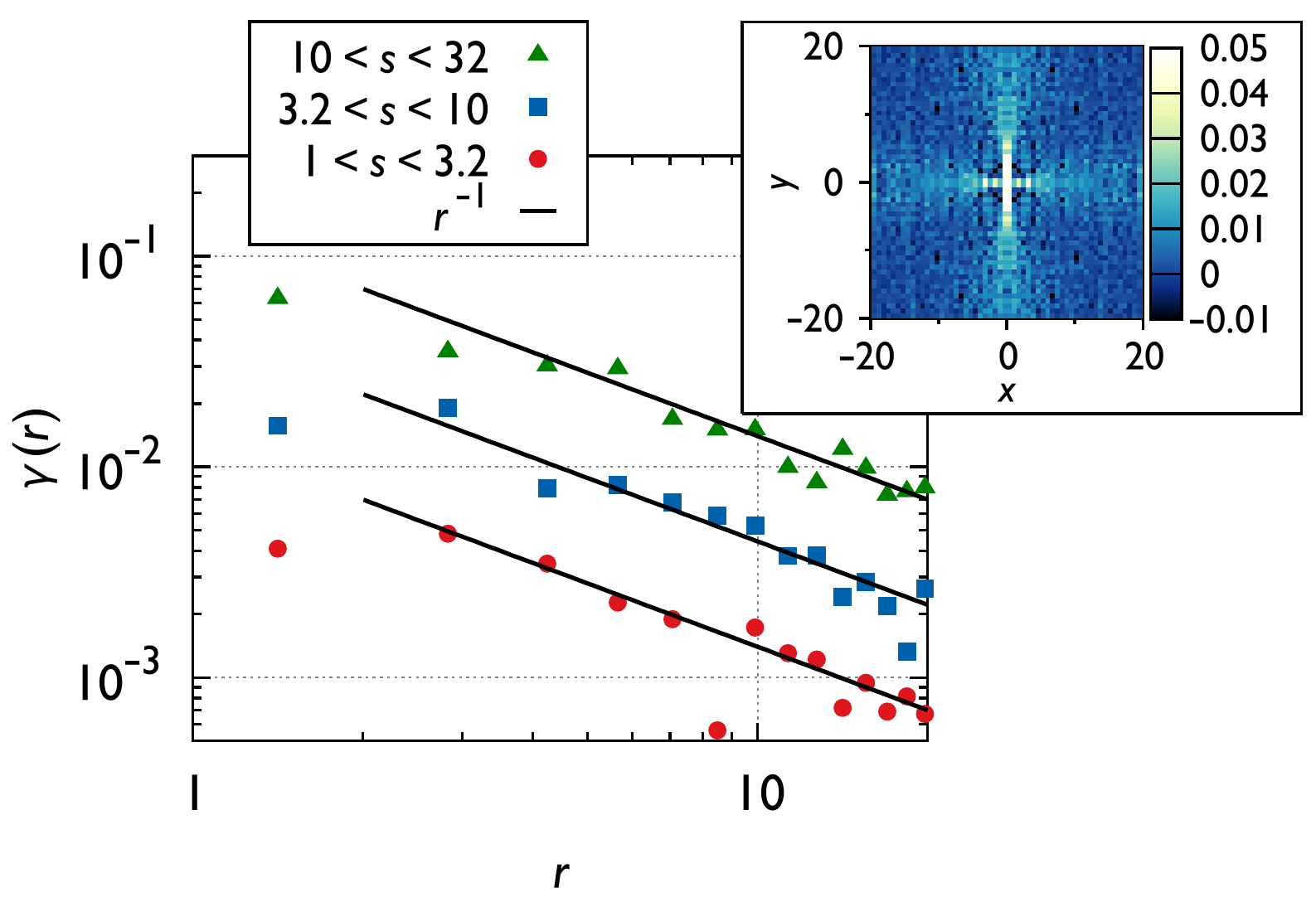}
\end{center}
\vspace*{-0.3cm}
\caption{(color online) Average spatial distribution of the avalanche plastic strain determined
for avalanches occurring at $\sigma_\text{ext} \approx 0.08$ in the CTD model for $N=2048$.
Main figure: Radial decay of the angle-averaged plastic strain for avalanches of different 
sizes.  Inset: Strain pattern $\gamma({\bf r})$ averaged over all avalanches. 
\label{fig:avalanche_structure}}
\end{figure}

Thus, our results demonstrate that there are system size effects at 
every stress level. To understand the nature of these size effects we note that
the spatial correlations of dislocation positions are short-ranged, with a 
correlation length $\xi$ of the order of the average dislocation spacing 
$\xi \approx 0.25$ (in the scaled units introduced above) \cite{ZAI-03,GRO-06} which 
defines the only internal length scale characterizing the dislocation structure. 
Given that the system sizes considered here are much larger, with $L$ ranging 
from 44$\xi$ ($N=128$) to 256$\xi$ ($N=4096$), the size effect we observe is not 
related to this microscopic length scale. Therefore, we consider instead
dynamic correlations in the motion of dislocations. To this end, we analyze 
the spatial structure of the avalanches in terms of the average spatial distribution 
of the plastic strain $\gamma({\bf r})$ produced during an avalanche and its angular 
average $\gamma(r)$ (these quantities relates to the avalanche size by 
$s=\int \gamma({\bf r}) d^2 r = \int 2 \pi r \gamma(r) d r$). To determine average values 
of these quantities, we shift the avalanche initiation points 
(taken to be the location of the fastest dislocation when $V_\text{th}$ is exceeded)
into the origin of a Cartesian coordinate system and then average the superimposed 
strain patterns over multiple avalanches. Figure \ref{fig:avalanche_structure} 
demonstrates that $\gamma({\bf r})$ exhibits a strongly anisotropic structure 
and decays slowly along the $x$ and $y$ axes. Averaged over all directions, the radial decay 
is of $1/r$ type regardless of the avalanche size $s$. This indicates that the 
long-range stress fields of the moving dislocations are not fully screened (contrary 
to what is observed in equilibrium \cite{GRO-06}), leading to highly 
non-local spreading of the avalanche activity. Thus, even at low stresses avalanches are 
influenced by the finite system size, naturally leading to an $L$-dependent 
slip avalanche cutoff. The fact that this cutoff diverges with increasing $L$ indicates
that in the thermodynamic limit the system is scale-free in the dynamic sense even for 
small applied stresses. Analogous conclusions can be drawn from an investigation of the 
velocity distributions of dislocations during various relaxation scenarios \cite{ISP-11}
which demonstrates that the distributions follow at all stresses a simple scaling relation 
indicating the absence of a time-scale in the system.

To conclude, we have established that the statistics of slip avalanches in simple $2d$ DDD models
is inconsistent with a depinning transition. Fundamental differences between the behavior 
of dislocation systems and the interface pinning/depinning scenario are manifested by the behavior 
of the cut-off of the avalanche size distribution which, rather than diverging at some critical stress
$\sigma_{\rm c}$, scales exponentially with stress but diverges with system size at every stress level. 
In addition, the avalanche exponent $\tau \approx 1.0$ is inconsistent with MFD. Rather than depinning,
a possible analogy for the observed behavior is provided by avalanches in glassy systems, such as 
mean-field spin glasses \cite{PAZ-99,DOU-10}, where an exponent $\tau=1$ and a size effect analogous to 
Eq.~(\ref{eq:cutoff_scaling}) have been observed. This is in line with other investigations which have 
shown that dislocation systems exhibit typical glassy properties such as slow relaxation 
\cite{ISP-11,CSI-09,CSI-05} and aging \cite{BAK-07}. 

While we have demonstrated that the equation "Yielding = Depinning" is not generally valid, it is 
important to note that real dislocation systems are composed of flexible lines moving in three dimensions 
and their behavior may differ from the present, highly idealized 2$d$ models. Therefore, both
$3d$ DDD simulation studies and experimental studies with large statistical samples are required
in order to understand the stress-resolved statistics of dislocation avalanches in 3$d$ and 
to settle the question regarding the fundamental nature of the yielding/jamming transition of 
dislocation systems.

{\bf Acknowledgments}. This work has been supported by the Hungarian Scientific and Research 
Fund (P.D.I.\ and I.G.,\ project nos.\ OTKA-PD-105256 and OTKA-K-105335); the European 
Commission (P.D.I.,\ project no.\ MC-CIG-321842); and the Academy of Finland through a 
Postdoctoral Researcher's Project (L.L.,\ project no.\ 139132), through an Academy Research
Fellowship (L.L.,\ project no.\ 268302), through the Centres of 
Excellence Program (project no.\ 251748), and via two travel grants (L.L.,\ P.D.I.,\ nos.\ 
261262 and 261521). The numerical simulations presented above were partly performed using 
computer resources within the Aalto University School of Science ``Science-IT'' project.
S.Z.\ is supported by the European Research Council, AdG2001-SIZEFFECTS and thanks the visiting 
professor program of Aalto University.

\end{document}